\documentclass[%
 aip,
 jmp,%
 amsmath,amssymb,
 reprint,%
]{revtex4-1}

\usepackage{graphicx}
\usepackage{subfigure}
\usepackage{hyperref}
\usepackage{xcolor}
\usepackage{units}
\hypersetup{colorlinks=true, linkcolor=darkblue,urlcolor=black, citecolor=darkgreen}
\definecolor{darkblue}{rgb}{0,0,0.3}
\definecolor{darkgreen}{rgb}{0,0.3,0}
\usepackage{dcolumn}
\usepackage{bm}

\begin{document}

\preprint{AIP/123-QED}

\title[]{Spin Currents injected electrically and thermally from highly spin polarized Co$_2$MnSi}

\author{Alexander Pfeiffer}
\affiliation{ 
Institut f\"ur Physik, Johannes Gutenberg-Universit\"at Mainz, 55099 Mainz, Germany
}

\author{Shaojie Hu}
\affiliation{
Research Center for Quantum Nano-Spin Sciences, Kyushu University, 6-10-1 Hakozaki, Fukuoka, 812-8581, Japan
}

\author{Robert M. Reeve}
\affiliation{ 
Institut f\"ur Physik, Johannes Gutenberg-Universit\"at Mainz, 55099 Mainz, Germany
}

\author{Alexander Kronenberg}
\affiliation{ 
Institut f\"ur Physik, Johannes Gutenberg-Universit\"at Mainz, 55099 Mainz, Germany
}

\author{Martin Jourdan}
\affiliation{ 
Institut f\"ur Physik, Johannes Gutenberg-Universit\"at Mainz, 55099 Mainz, Germany
}

\author{Takashi Kimura}
\affiliation{
Research Center for Quantum Nano-Spin Sciences, Kyushu University, 6-10-1 Hakozaki, Fukuoka, 812-8581, Japan
}
\affiliation{
Department of Physics, Kyushu University, 6-10-1 Hakozaki, Fukuoka, 812-8581, Japan
}

\author{Mathias Kl\"aui}
\email{klaeui@uni-mainz.de}
\affiliation{ 
Institut f\"ur Physik, Johannes Gutenberg-Universit\"at Mainz, 55099 Mainz, Germany
}

\date{\today}

\begin{abstract}
We demonstrate the injection and detection of electrically and thermally generated spin currents probed in Co$_2$MnSi/Cu lateral spin valves.  Devices with different electrode separations are patterned to measure the non-local signal as a function of the electrode spacing and we determine a relatively high effective spin polarization  $\alpha$ of  Co$_2$MnSi to be 0.63 and the spin diffusion length of Cu to be \unit[500]{nm} at room temperature. The electrically generated non-local signal is measured as a function of temperature and a maximum signal is observed for a temperature of \unit[80]{K}. The thermally generated non-local signal is measured as a function of current density and temperature in a second harmonic measurement detection scheme. We find different temperature dependences for the  electrically and thermally generated non-local signals, which allows us to conclude that the temperature dependence of the signals is not just dominated by  the transport in the Cu wire, but that there is a crucial contribution from the different generation mechanisms, which has been largely disregarded to date.
\end{abstract}

\maketitle

Recently pure spin currents have been receiving a great deal of attention as an exciting and efficient new means of manipulating the magnetic state of a device, while potentially reducing disadvantageous Joule heating and Oersted field effects at the position of the ferromagnet that is manipulated. Non-local spin valves, consisting of two spatially separated magnetic nano-structures bridged by a nonmagnetic channel, have been intensively studied as an easy possibility to generate and detect pure spin currents via spin injection from the injector into the conduit.~\cite{Jedema,Johnson} This leads to a spin accumulation which diffuses away from the injection point and comprises a pure diffusive spin current in the direction of the second ferromagnetic electrode, where the spin current can be detected and manipulate the local magnetization due to the spin transfer torque.~\cite{Yang, Ilgaz} A large efficiency for a domain wall displacement assisted by a pure spin current~\cite{Ilgaz} and even  pure spin current induced domain wall displacement was reported.~\cite{Motzko} Furthermore, non-local spin valves have recently started to be intensely investigated as a geometry for future magnetic read-head devices.~\cite{Yamada} Scientifically,  the non-local technique allows for the determination of key parameters of the spin transport, namely the spin polarization $\alpha$ of the ferromagnetic and the spin diffusion length of the nonmagnetic material~\cite{Yang} and based on these parameters the ratio between spin and charge current.~\cite{Otani} To generate larger spin currents and improve the efficiency of magnetization manipulation  but also to obtain larger signals for use in read-heads,  recent studies used Heusler  based ferromagnetic electrodes and found large spin signals.~\cite{Takahashi,Bridoux,Hamaya,KimuraCFS,Kento,Hu,Yang2}\\
 One promising material for non-local spin valves is  the Heusler compound Co$_2$MnSi for which recently \unit[100]{\%} spin polarization at room temperature was observed.~\cite{Jourdan} \\
In addition to these electrically generated spin currents, it has recently been demonstrated that it is also possible to generate thermal spin currents in non-local spin valves via the spin dependent Seebeck effect due to the thermal gradient that is established at the injector-conduit interface as a result of Joule heating.~\cite{Schlachter,Bakker,Erekhinsky,Hu,Kento} While potentially very useful for instance for waste-heat recovery, for this  generation method, spin current efficiencies currently tend to be lower than the electrical injection method, calling for improvements. Hu et al.   made recent progress in this regard by employing a CoFeAl thermal spin current injector electrode.~\cite{Hu} The resulting dramatic improvements to the spin injection efficiency were partly attributed to the large spin-dependent Seebeck coefficient which arises from the change in sign in the Seebeck coefficient for spin up and down electrons due to the large difference in DOS for the different spin-subbands at the Fermi level and partly attributed to the low spin  resistance leading to reduced backflow  spin absorption.~\cite{Kimura3}\\
In the present work we study  Co$_2$MnSi/Cu non-local spin valves. By measuring the electrically generated non-local signal as a function of the electrode separation we are able to extract the key parameters for our devices at room temperature finding a spin diffusion length of \unit[500]{nm} in the Cu conduit and a  spin polarization of 0.63 at room temperature. The reduction of the measured spin polarization as compared to the previously measured intrinsic spin polarization of the Heusler material is attributed to the lower interface spin polarization, highlighting the importance of the interface properties on the measured signals.  We compare the electrically  generated non-local signals as a function of temperature with thermally generated signals in the same sample. These measurements show a significantly different temperature behaviour, which can be explained by different temperature dependencies of the injection mechanisms at the Co$_2$MnSi/Cu interface. Such contributions have been largely overlooked in previous studies, where mostly spin relaxation mechanisms in the non-magnetic material are proposed to account for the temperature evolution of the signals,  which  are clearly not sufficient to explain our results.~\cite{Kimura2,Villamor,OBrien,Motzko2}

\begin{figure}
\subfigure[]
{\includegraphics[width=0.14 \textwidth]{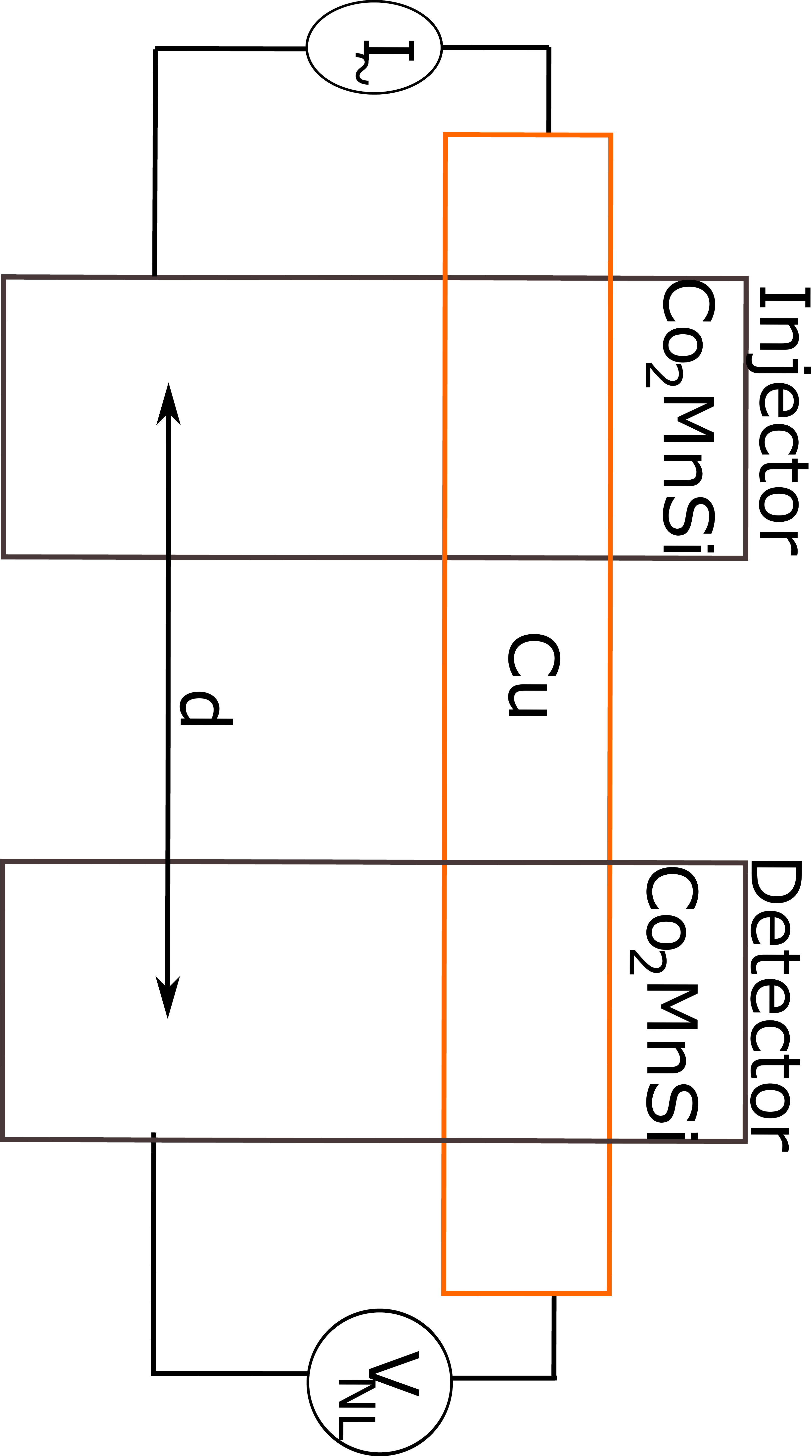}}
\subfigure[]
{\includegraphics[width=0.22 \textwidth]{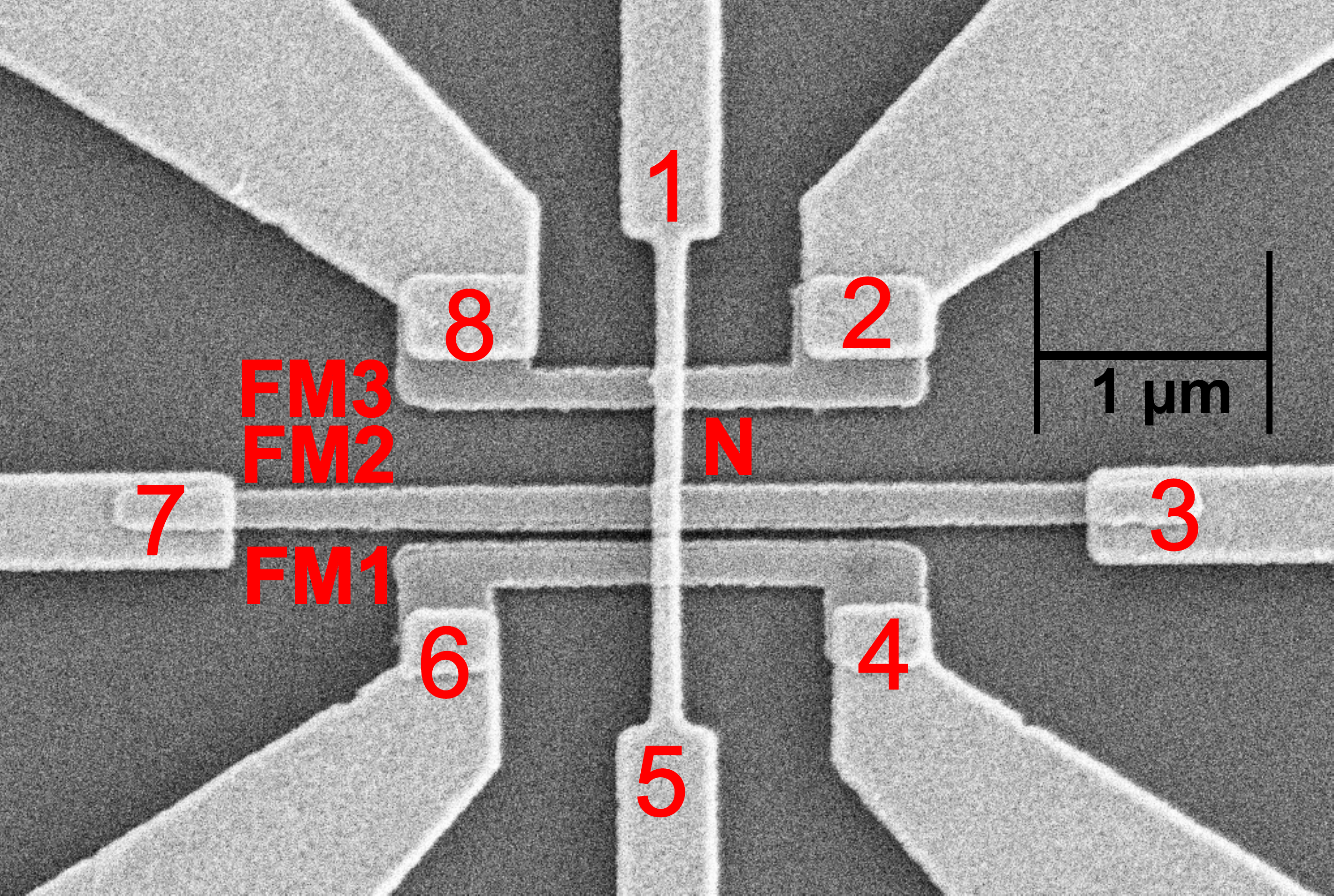}}
\caption{(a)  Schematic illustration of a non-local spin valve consisting of two nanowires made of Co$_2$MnSi bridged by a Cu nanowire.\\
(b) High resolution scanning electron microscope image with numbered contacts and the nano wires labeled FM1, FM2, FM3 and N. Three nanowires provide the possibility to measure two non-local signals with one device using FM2 as  detector and FM1 or FM3 as injector.}
\label{SEM}
\end{figure}

A schematic depiction and a scanning electron microscope image of the patterned nanowires is shown in \autoref{SEM}. A \unit[47]{nm} thick Co$_2$MnSi and \unit[42]{nm} thick Ag  thin film is grown via rf sputtering and then nano-structured to obtain about \unit[200]{nm} wide ferromagnetic electrodes with a \unit[200]{nm} thick and \unit[170]{nm} wide Cu nanowire on top by two-step lift-off electron beam lithography. We have confirmed the \unit[100]{\%} intrinsic spin polarization of the grown Co$_2$MnSi as described in~\cite{Jourdan}. Before depositing the Cu nanowire, 30 second Ar$^+$-ion milling is performed to provide clean interfaces. Different devices, fabricated from the same film and made in the same batch, were measured  with a range in electrode separations from \unit[200]{nm} to \unit[1000]{nm}.  In order to characterize the devices we perform two sets of measurements. Firstly, for the study of spin currents via electrical spin-injection we apply a small \unit[220]{$\mu$A} alternating current between the injector and the end of the Cu wire which acts as the non-magnetic (N)  spin current conduit  (contact 6 and contact 5 as shown in \autoref{SEM}). This generates a spin accumulation which diffuses as a pure spin current towards the interface between the Cu wire  and the detector. 
If the spin resistance is low,~\cite{Kimura3} the pure spin current is then strongly  absorbed in the detector. Depending on the relative orientation of the magnetization of the two ferromagnetic electrodes, a high non-local voltage for a parallel alignment and a low  non-local voltage for an antiparallel alignment of the magnetization of the nanowires is measured between the detector and the conduit (contact 7 and contact 1) in the first harmonic measurement via a standard lock-in technique.
As a second effect a stronger applied current leads to significant Joule heating of the FM1/N interface. The resulting temperature gradient between the hot junction and the cold sample environment gives rise to a thermally generated spin current due to the spin dependent Seebeck effect.~\cite{Schlachter,Hu,Kento} Since the measured voltage scales quadratically with the applied charge current due to Joule's law, we measure the thermally generated non-local voltage as the second harmonic signal with the same probe configuration as for the electrically generated signal. Temperature dependent measurements are carried out for certain temperatures between 5 K and 300 K in the same sample, where we keep the applied charge current and the temperature constant during the measurement of the non-local signal. For a quantitative comparison in the case of electrically generated spin currents, we define a non-local resistance via $V_{\text{NL}}/I_{\text{ac}}$ where $V_{\text{NL}}$ is the measured non-local signal and $I_{\text{ac}}$ the applied charge current of about \unit[220]{$\mu$A}.

\begin{figure}
\subfigure[]
{\includegraphics[width=0.40 \textwidth]{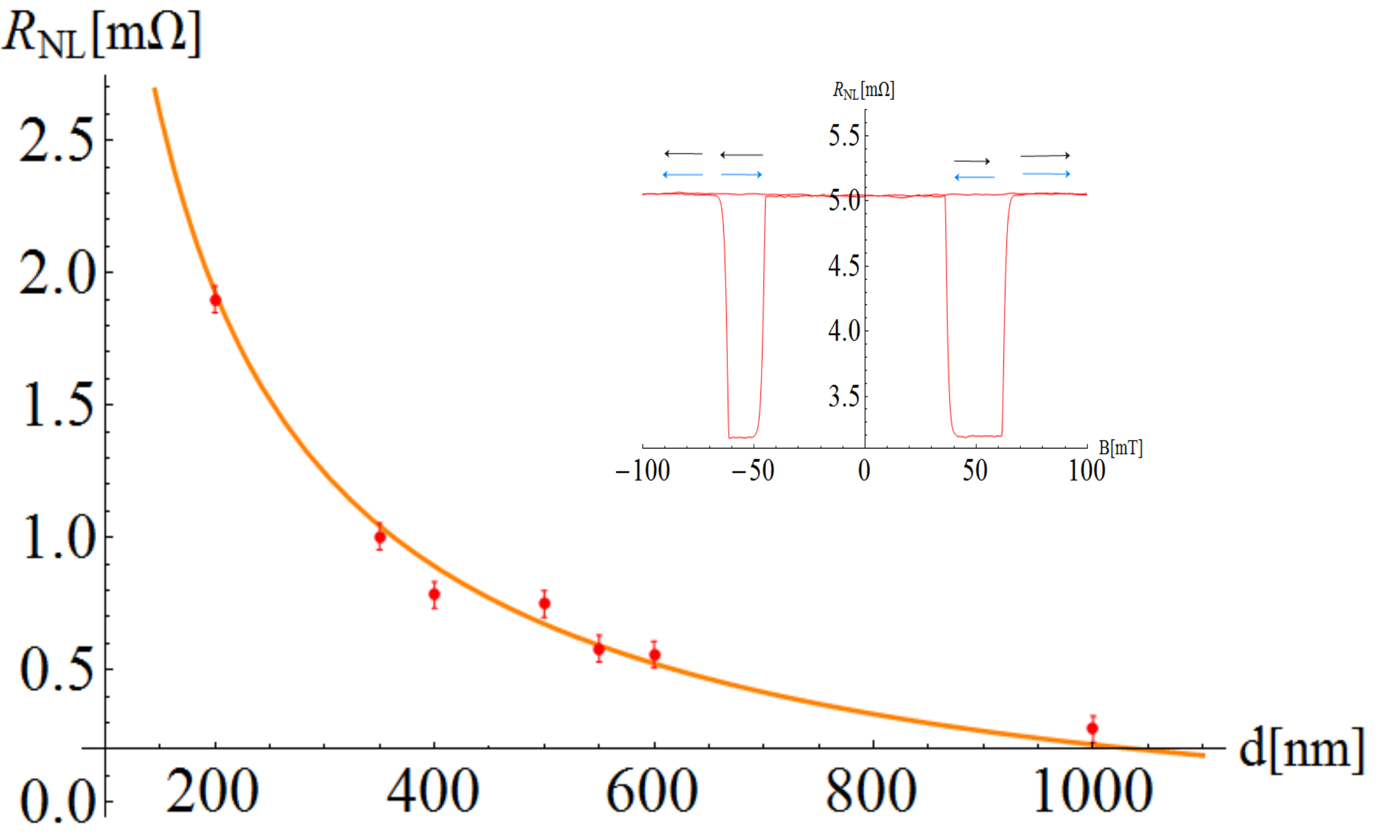}} 
\subfigure[]
{\includegraphics[width=0.40 \textwidth]{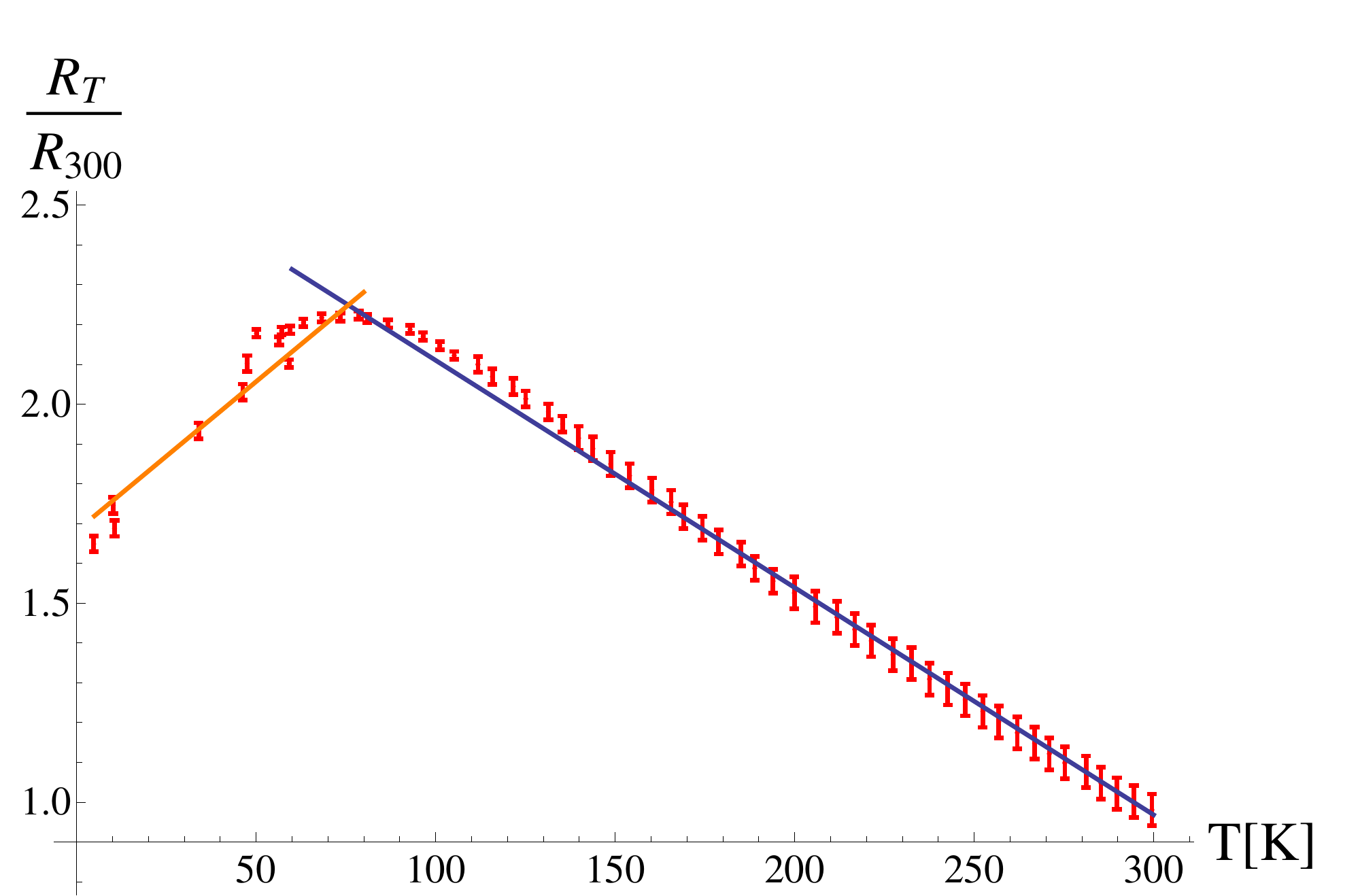}}
\caption{
(a)  Observed non-local signal as a function of the electrode distance measured at room temperature and the corresponding fit based on \autoref{EleFormel}. \\
(b) Normalized non-local signal as a function of temperature where we observe a maximum signal at a temperature of \unit[80]{K} and a reduction of the signal of \unit[40]{\%} between \unit[5]{K} and \unit[80]{K}.}
\label{NonLokalEle}
\end{figure}

A non-local curve with its characteristic two states while sweeping the field along the easy axes of the wires is shown inside in \autoref{NonLokalEle}\textcolor{darkblue}{(a)}.
Furthermore  we present in this figure the non-local resistance as a function of the electrode distance where the red dots are the measured data points and the orange curve is a fit using an analytical solution~\cite{Otani,MaekawaTaka}  of the spin diffusion problem in 1-D:
\begin{align}
\Delta R_{\text{NL}}(\text{d})= \frac{\alpha^2 R_\text{{F}}^2 R_\text{{N}}}{\exp{( d/ \lambda_N)}[ 2 R_{\text{N}} R_{\text{F}}+ 2 R_{\text{F}}^2]+ R_\text{N}^2 \sinh( d/ \lambda_N).}
\label{EleFormel}
\end{align}  

In this formula,  $\alpha$ is the effective spin polarization of  Co$_2$MnSi and $\lambda_N$ the spin diffusion length of Cu. $R_\text{N}$ and $R_\text{F}$ are the spin resistances of the  nonmagnetic and ferromagnetic  material which are defined as $R_\text{S,i}= 2 \rho_i \cdot \lambda_i/(S(1- \alpha_i)^2)$ with $\rho$ as the resistivity and $S$ as the effective cross sectional area where the spin current flows. $\lambda_F$ is assumed to be \unit[1]{nm}, the  average effective cross section $S$ is determined to  be \unit[33000]{nm$^2$} while the resistivities are  measured  as \unit[3.2$\cdot$10$^{-7}$]{$\Omega \cdot$m} for Co$_2$MnSi and \unit[1.8$\cdot$10$^{-8}$]{$\Omega\cdot$m} for Cu.   
We determine the spin diffusion length $\lambda_N$ of Cu to be \unit[500]{nm} which is in agreement with other reports~\cite{Kimura, Zou, Casanova, Villamor} and determine $\alpha$ to be 0.63 at room temperature. While higher than for typical 3d metals, this spin polarization value is lower than the recently reported intrinsic spin polarization of Co$_2$MnSi~\cite{Jourdan} which implies significant spin relaxation occurs  when the current is injected across the Co$_2$MnSi/Cu interface. For transparent contacts sources of such spin relaxation can include absorption of the spin accumulation by the ferromagnet due to a low spin resistance,~\cite{Kimura2} increased backscattering of electrons from the start of the conduit and spin flip scattering of electrons as they cross the interface or at impurities in the vicinity of the interface.~\cite{Erekhinsky2} However, for a perfect spin polarized material the spin relaxation within the ferromagnet should be suppressed.~\cite{KimuraCFS} In our case  the junction resistances are relatively high for an all metallic system (on the order of \unit[5]{$\Omega$}), which means that indeed spin flip at the interface can occur. In this case we are probing the interfacial spin polarization in the measurement which can be expected to be lower than the intrinsic spin polarization.~\cite{Brass}  To circumvent this problem in the future, dedicated materials combinations have to be developed where the interfacial polarization of the spins that enter the non-magnetic spin conduit is \unit[100]{\%} , which can possibly be achieved by band-structure matching and using an all-Heusler stack, which is though beyond the scope of the current work.

In \autoref{NonLokalEle}\textcolor{darkblue}{(b)} we show the non-local signal as a function of temperature in the range of \unit[5]{K} and \unit[300]{K} for the \unit[350]{nm} separation device. Initially on reducing the temperature the signal displays a monotonic increase, as expected from the Elliott-Yafet spin relaxation mechanism in the case of a monotonic decrease in resistivity on cooling.~\cite{Elliott} However, around \unit[80]{K} a maximum in the signal is observed and at lower temperatures the signal decreases significantly. Whilst an obvious low temperature decrease in signal is not apparent in the recently studied Heusler non-local spin valves,~\cite{Hamaya,KimuraCFS} the occurrence of a maximum of the non-local signal at a certain temperature is well studied in literature.~\cite{Kimura2,Villamor,OBrien,Motzko2}  The physical origin for this non monotonic behaviour is still an open question, with a variety of explanations having been put forward including surface scattering in the conduit, surface oxidation, grain boundaries, magnetic impurities and a manifestation of a Kondo effect.~\cite{Kimura2,Villamor,OBrien,Motzko2}

To shed light onto the origin of the temperature dependence, we compare further below the electrically injected spin current to a thermally generated spin current. If the temperature dependence is dominated by an effect of the spin current propagation in the spin current conduit, such as surface oxidation of the conduit, different injection methods should yield the same temperature dependence.

\begin{figure}

\subfigure[]
{\includegraphics[width=0.40 \textwidth]{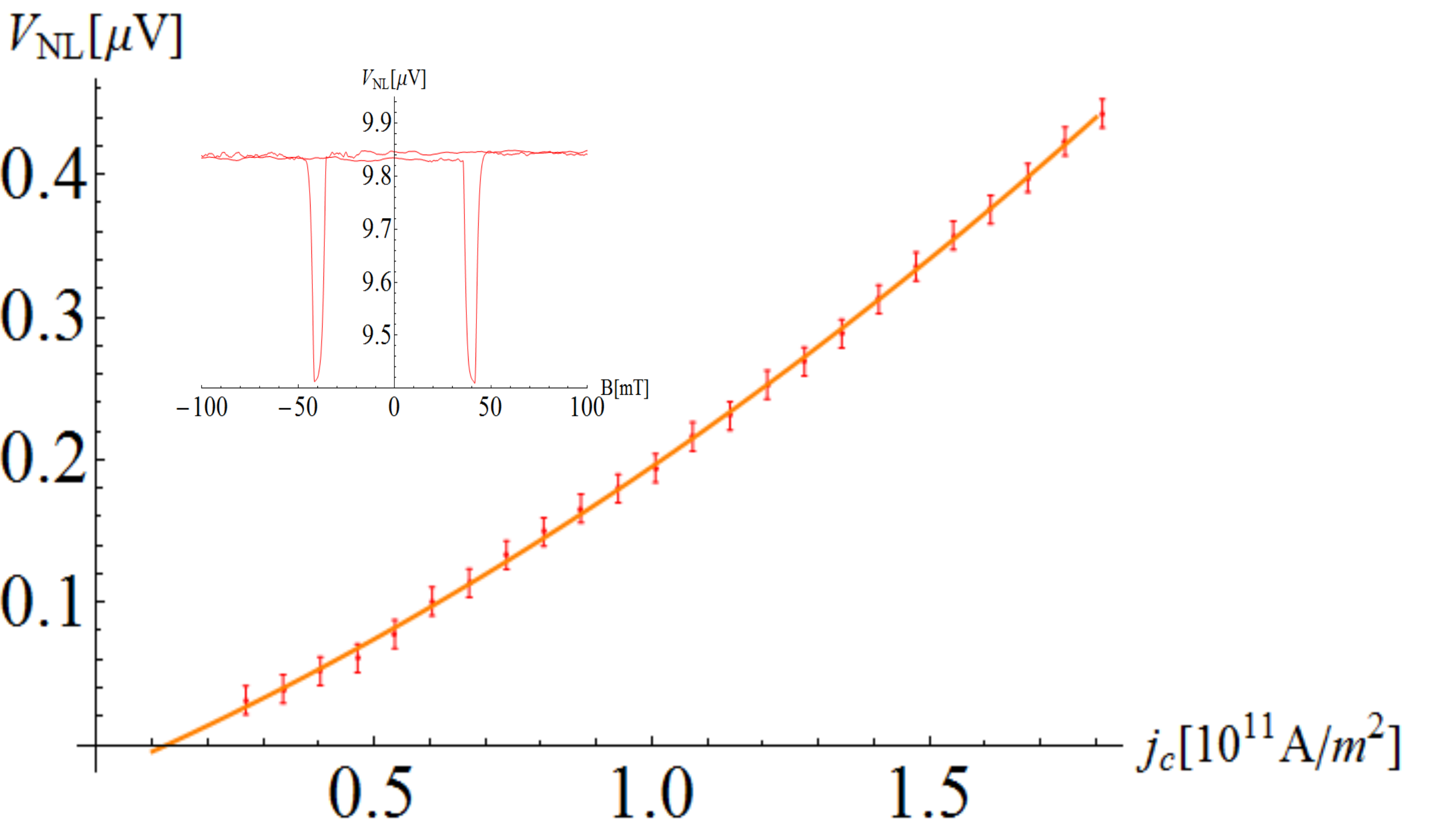}} 
\subfigure[]
{\includegraphics[width=0.40 \textwidth]{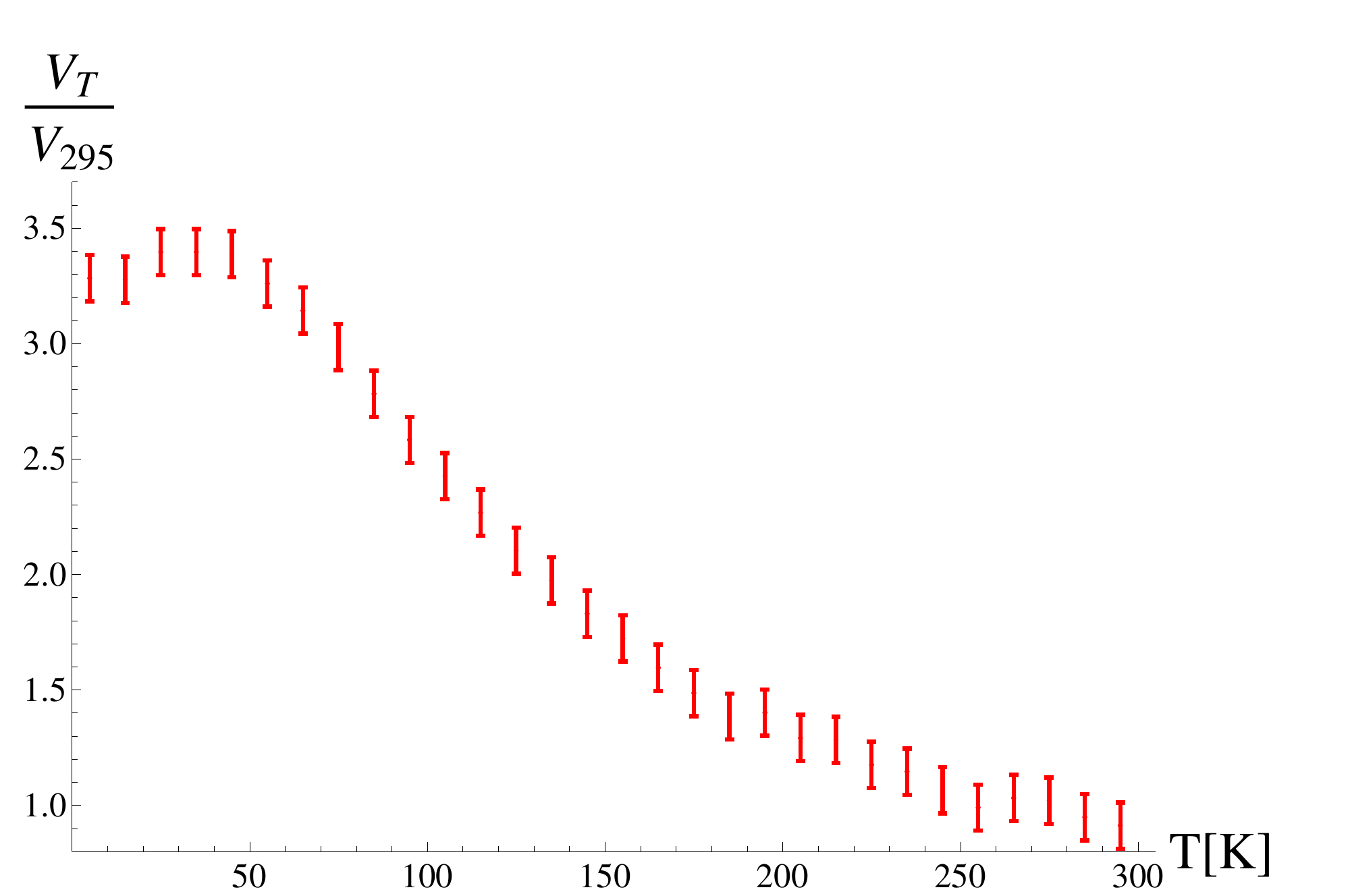}}
\caption{
(a) Thermal signal as a function of the current density. We find the expected quadratic behaviour as a consequence of Joule heating  showing that the spin current is indeed due to thermal spin injection. \\
(c) Normalized thermal signal as a  function of temperature with a wide increase of the signal between \unit[300]{K} and \unit[70]{K} and a constant signal at very low temperatures.}
\label{NonLokalThermal}
\end{figure}

So to check this, a thermally generated non-local curve, measured with an electrode distance of \unit[350]{nm} and  an applied current density of \unit[1.8$\cdot$10$^{11}$]{A/m$^2$} while sweeping the field, is shown inside in \autoref{NonLokalThermal}\textcolor{darkblue} {(a)} in addition with the observed thermal signal as a function of current density. A similar switching behaviour to the electrically generated signals is observed. Compared to the electrically generated non-local curve, the voltage drops are observed for a narrower field range instead of a wide field plateau for the  antiparallel state  for the electrically generated spin currents with less thermal excitation. Nevertheless we can identify the  voltage drop with the antiparallel alignment and average over 50 field value  data points in the lower state resulting in a low statistical error for the thermal signal as the difference between parallel and antiparallel alignment  allowing us to extract the amplitude of the signal with high confidence.

In \autoref{NonLokalThermal}\textcolor{darkblue}{(b)}, we show the normalized  thermal generated non-local signal. Please note that the applied current density is \unit[1.2$\cdot$10$^{11}$]{A/m$^2$} for these measurements, which is low enough to prevent any significant increase of the resistance between contact 6 and contact 5, meaning that the cryostat temperature is indeed the sample temperature. The thermally generated signal shows a very different low temperature behaviour than the electrically generated one. We do not observe a maximum but a constant signal at temperatures between \unit[5]{K} and \unit[65]{K} and also the size of the relative signal differs significantly. We measure in the temperature range between \unit[295]{K} and \unit[75]{K} an increase of the signal of \unit[350]{\%} while for the electrically generated signal we just observe an increase of \unit[220]{\%}.

These observations motivate a comparison between both types of spin currents, their temperature dependence and their physical origin.\\
In both cases, the resulting signal depends both on the magnitude of the spin accumulation initially generated in the conduit and the decay of that spin accumulation as it approaches the detector. At higher temperatures,  the spin relaxation in the conduit is expected to be dominated by the Elliott-Yafet spin relaxation mechanism,~\cite{Elliott} since additional proposed contributions to spin relaxation such as surface scattering along the conduit length,~\cite{Kimura2,Villamor,Motzko2} or the Kondo effect in the vicinity of the interface~\cite{OBrien} are only expected to potentially play a role at low temperatures. In the Elliott-Yafet mechanism,  the spin-flip scattering rate is proportional to the momentum scattering rate and therefore should scale with the resistivity of the conduit identically, regardless of the initial generation mechanism.~\cite{Elliott} This explains the general trend for the increase in signal with decreasing temperature in both cases in the temperature range between \unit[300]{K} and about \unit[80]{K}. In the case of the thermal spin injection, the temperature gradient in the vicinity of the injector will complicate the picture since the spin diffusion length may not be uniform along the length of the conduit. This may partly explain why above \unit[100]{K} the electrical signal is well described by a linear function whereas the thermal spin signal has a changing slope.

At lower temperatures a qualitative difference is observed between the two curves. For temperatures below \unit[80]{K} we observe a maximum in the electrically generated non-local signal and for lower temperatures a strong reduction of \unit[40]{\%} to the maximum signal whilst in the case of thermally generated spin currents, we observe a plateau between \unit[65]{K} and \unit[5]{K}. In order to explain the non monotonic behaviour in this regime, additional temperature dependent spin relaxation pathways are usually invoked in part or all of the conduit~\cite{Motzko2,Kimura2,OBrien}. However, if these were the only dominant effects we would expect no difference for the two cases. We therefore deduce that the two different temperature dependences must include effects from the different generation mechanisms and cannot purely originate from the temperature dependence of the spin transport in the spin current conduit. This is a key result since such effects are largely disregarded in the existing literature where the temperature dependence was explained only by changes in the transport properties of the nonmagnetic channel.~\cite{Motzko2,Kimura2,Villamor,OBrien} Yet for both spin current sources a temperature dependence of the generated spin current may be expected with changes to the polarization of electrically injected carriers and variations in the temperature gradients in the device, in addition to an evolution in the junction resistance.

In conclusion, we studied electrically and thermally generated spin currents in Co$_2$MnSi/Cu spin valve structures as a function of temperature. In the case of electrical injection we determine
the spin transport parameters of the devices by measuring the non-local signal as a function of electrode separation. From our fit we find the effective spin polarization of Co$_2$MnSi and the room temperature spin diffusion length of Cu, which are determined to be \unit[63]{\%} and \unit[500]{nm} respectively. While this is a relatively high spin polarization, a significant contribution to spin relaxation due to the interface properties is suggested, calling for future care in interface engineering and choice of materials combinations to achieve the desired maximum spin accumulation in the conduit.

 By  direct comparison between the electrically and thermally generated signals as a function of temperature in the same device, we study the temperature dependence of the injection and spin transport. Since the spin transport in the nonmagnetic channel is diffusive for both types of spin currents~\cite{Hu} the same temperature behaviour would be expected if effects in the conduit dominated both measurements. However, a qualitative difference in behaviour is observed, demonstrating that in contrast to most existing studies, the temperature dependence of the spin current generation mechanism must be taken into account and will have a decisive effect on the device behaviour.

\vspace*{0.5cm}

This work was funded by the German
Ministry for Education and Science (BMBF), the German Research
Foundation (DFG), the Graduate School Material Science in Mainz
(DFG/GSC 266), EU 7th Framework Programme (WALL FP7-PEOPLE-2013-ITN 608031; MAGWIRE, FP7-ICT-2009-5), the European Research Council through the Starting Independent
Researcher Grant MASPIC (ERC-2007-StG 208162), DFG SpinCat SPP1538 and the Research Center of Innovative and
Emerging Materials at Johannes Gutenberg University (CINEMA).
A. Pfeiffer and M. Kl\"aui thank the participants of SpinMechanics 2 Workshop for valuable discussions and are grateful for financial support from the German Academic Exchange Service (DAAD) via the SpinNet Program (56268455).

\bibliography{Pfeiffer1}

\begin{thebibliography}{30}%
\makeatletter
\providecommand \@ifxundefined [1]{%
 \@ifx{#1\undefined}
}%
\providecommand \@ifnum [1]{%
 \ifnum #1\expandafter \@firstoftwo
 \else \expandafter \@secondoftwo
 \fi
}%
\providecommand \@ifx [1]{%
 \ifx #1\expandafter \@firstoftwo
 \else \expandafter \@secondoftwo
 \fi
}%
\providecommand \natexlab [1]{#1}%
\providecommand \enquote  [1]{``#1''}%
\providecommand \bibnamefont  [1]{#1}%
\providecommand \bibfnamefont [1]{#1}%
\providecommand \citenamefont [1]{#1}%
\providecommand \href@noop [0]{\@secondoftwo}%
\providecommand \href [0]{\begingroup \@sanitize@url \@href}%
\providecommand \@href[1]{\@@startlink{#1}\@@href}%
\providecommand \@@href[1]{\endgroup#1\@@endlink}%
\providecommand \@sanitize@url [0]{\catcode `\\12\catcode `\$12\catcode
  `\&12\catcode `\#12\catcode `\^12\catcode `\_12\catcode `\%12\relax}%
\providecommand \@@startlink[1]{}%
\providecommand \@@endlink[0]{}%
\providecommand \url  [0]{\begingroup\@sanitize@url \@url }%
\providecommand \@url [1]{\endgroup\@href {#1}{\urlprefix }}%
\providecommand \urlprefix  [0]{URL }%
\providecommand \Eprint [0]{\href }%
\providecommand \doibase [0]{http://dx.doi.org/}%
\providecommand \selectlanguage [0]{\@gobble}%
\providecommand \bibinfo  [0]{\@secondoftwo}%
\providecommand \bibfield  [0]{\@secondoftwo}%
\providecommand \translation [1]{[#1]}%
\providecommand \BibitemOpen [0]{}%
\providecommand \bibitemStop [0]{}%
\providecommand \bibitemNoStop [0]{.\EOS\space}%
\providecommand \EOS [0]{\spacefactor3000\relax}%
\providecommand \BibitemShut  [1]{\csname bibitem#1\endcsname}%
\let\auto@bib@innerbib\@empty
\bibitem [{\citenamefont {Jedema}, \citenamefont {Filip},\ and\ \citenamefont
  {van Wees}(2001)}]{Jedema}%
  \BibitemOpen
  \bibfield  {author} {\bibinfo {author} {\bibfnamefont {F.~J.}\ \bibnamefont
  {Jedema}}, \bibinfo {author} {\bibfnamefont {A.~T.}\ \bibnamefont {Filip}}, \
  and\ \bibinfo {author} {\bibfnamefont {B.~J.}\ \bibnamefont {van Wees}},\
  }\href@noop {} {\bibfield  {journal} {\bibinfo  {journal} {Nature}\ }\textbf
  {\bibinfo {volume} {410}},\ \bibinfo {pages} {345--348} (\bibinfo {year}
  {2001})}\BibitemShut {NoStop}%
\bibitem [{\citenamefont {Johnson}\ and\ \citenamefont
  {Silsbee}(1988)}]{Johnson}%
  \BibitemOpen
  \bibfield  {author} {\bibinfo {author} {\bibfnamefont {M.}~\bibnamefont
  {Johnson}}\ and\ \bibinfo {author} {\bibfnamefont {R.~H.}\ \bibnamefont
  {Silsbee}},\ }\href {\doibase 10.1103/PhysRevB.37.5312} {\bibfield  {journal}
  {\bibinfo  {journal} {Phys. Rev. B}\ }\textbf {\bibinfo {volume} {37}},\
  \bibinfo {pages} {5312--5325} (\bibinfo {year} {1988})}\BibitemShut {NoStop}%
\bibitem [{\citenamefont {Yang}, \citenamefont {Kimura},\ and\ \citenamefont
  {Y.}(2008)}]{Yang}%
  \BibitemOpen
  \bibfield  {author} {\bibinfo {author} {\bibfnamefont {T.}~\bibnamefont
  {Yang}}, \bibinfo {author} {\bibfnamefont {T.}~\bibnamefont {Kimura}}, \ and\
  \bibinfo {author} {\bibfnamefont {O.}~\bibnamefont {Y.}},\ }\href@noop {}
  {\bibfield  {journal} {\bibinfo  {journal} {Nature Phys.}\ }\textbf {\bibinfo
  {volume} {4}},\ \bibinfo {pages} {851--854} (\bibinfo {year}
  {2008})}\BibitemShut {NoStop}%
\bibitem [{\citenamefont {Ilgaz}\ \emph {et~al.}(2010)\citenamefont {Ilgaz},
  \citenamefont {Nievendick}, \citenamefont {Heyne}, \citenamefont {Backes},
  \citenamefont {Rhensius}, \citenamefont {Moore}, \citenamefont {Ni\~no},
  \citenamefont {Locatelli}, \citenamefont {Mente\ifmmode~\mbox{\c{s}}\else
  \c{s}\fi{}}, \citenamefont {v.~Schmidsfeld}, \citenamefont {v.~Bieren},
  \citenamefont {Krzyk}, \citenamefont {Heyderman},\ and\ \citenamefont
  {Kl\"aui}}]{Ilgaz}%
  \BibitemOpen
  \bibfield  {author} {\bibinfo {author} {\bibfnamefont {D.}~\bibnamefont
  {Ilgaz}}, \bibinfo {author} {\bibfnamefont {J.}~\bibnamefont {Nievendick}},
  \bibinfo {author} {\bibfnamefont {L.}~\bibnamefont {Heyne}}, \bibinfo
  {author} {\bibfnamefont {D.}~\bibnamefont {Backes}}, \bibinfo {author}
  {\bibfnamefont {J.}~\bibnamefont {Rhensius}}, \bibinfo {author}
  {\bibfnamefont {T.~A.}\ \bibnamefont {Moore}}, \bibinfo {author}
  {\bibfnamefont {M.~A.}\ \bibnamefont {Ni\~no}}, \bibinfo {author}
  {\bibfnamefont {A.}~\bibnamefont {Locatelli}}, \bibinfo {author}
  {\bibfnamefont {T.~O.}\ \bibnamefont {Mente\ifmmode~\mbox{\c{s}}\else
  \c{s}\fi{}}}, \bibinfo {author} {\bibfnamefont {A.}~\bibnamefont
  {v.~Schmidsfeld}}, \bibinfo {author} {\bibfnamefont {A.}~\bibnamefont
  {v.~Bieren}}, \bibinfo {author} {\bibfnamefont {S.}~\bibnamefont {Krzyk}},
  \bibinfo {author} {\bibfnamefont {L.~J.}\ \bibnamefont {Heyderman}}, \ and\
  \bibinfo {author} {\bibfnamefont {M.}~\bibnamefont {Kl\"aui}},\ }\href
  {\doibase 10.1103/PhysRevLett.105.076601} {\bibfield  {journal} {\bibinfo
  {journal} {Phys. Rev. Lett.}\ }\textbf {\bibinfo {volume} {105}},\ \bibinfo
  {pages} {076601} (\bibinfo {year} {2010})}\BibitemShut {NoStop}%
\bibitem [{\citenamefont {Motzko}\ \emph {et~al.}(2013)\citenamefont {Motzko},
  \citenamefont {Burkhardt}, \citenamefont {Richter}, \citenamefont {Reeve},
  \citenamefont {Laczkowski}, \citenamefont {Savero~Torres}, \citenamefont
  {Vila}, \citenamefont {Attan\'e},\ and\ \citenamefont {Kl\"aui}}]{Motzko}%
  \BibitemOpen
  \bibfield  {author} {\bibinfo {author} {\bibfnamefont {N.}~\bibnamefont
  {Motzko}}, \bibinfo {author} {\bibfnamefont {B.}~\bibnamefont {Burkhardt}},
  \bibinfo {author} {\bibfnamefont {N.}~\bibnamefont {Richter}}, \bibinfo
  {author} {\bibfnamefont {R.}~\bibnamefont {Reeve}}, \bibinfo {author}
  {\bibfnamefont {P.}~\bibnamefont {Laczkowski}}, \bibinfo {author}
  {\bibfnamefont {W.}~\bibnamefont {Savero~Torres}}, \bibinfo {author}
  {\bibfnamefont {L.}~\bibnamefont {Vila}}, \bibinfo {author} {\bibfnamefont
  {J.-P.}\ \bibnamefont {Attan\'e}}, \ and\ \bibinfo {author} {\bibfnamefont
  {M.}~\bibnamefont {Kl\"aui}},\ }\href {\doibase 10.1103/PhysRevB.88.214405}
  {\bibfield  {journal} {\bibinfo  {journal} {Phys. Rev. B}\ }\textbf {\bibinfo
  {volume} {88}},\ \bibinfo {pages} {214405} (\bibinfo {year}
  {2013})}\BibitemShut {NoStop}%
\bibitem [{\citenamefont {Yamada}\ \emph {et~al.}(2013)\citenamefont {Yamada},
  \citenamefont {Sato}, \citenamefont {Yoshida}, \citenamefont {Sato},
  \citenamefont {Meguro},\ and\ \citenamefont {Ogawa}}]{Yamada}%
  \BibitemOpen
  \bibfield  {author} {\bibinfo {author} {\bibfnamefont {M.}~\bibnamefont
  {Yamada}}, \bibinfo {author} {\bibfnamefont {D.}~\bibnamefont {Sato}},
  \bibinfo {author} {\bibfnamefont {N.}~\bibnamefont {Yoshida}}, \bibinfo
  {author} {\bibfnamefont {M.}~\bibnamefont {Sato}}, \bibinfo {author}
  {\bibfnamefont {K.}~\bibnamefont {Meguro}}, \ and\ \bibinfo {author}
  {\bibfnamefont {S.}~\bibnamefont {Ogawa}},\ }\bibfield  {title} {\enquote
  {\bibinfo {title} {Scalability of {S}pin {A}ccumulation {S}ensor},}\ }\href
  {\doibase 10.1109/TMAG.2012.2226871} {\bibfield  {journal} {\bibinfo
  {journal} {Magnetics, IEEE Transactions on}\ }\textbf {\bibinfo {volume}
  {49}},\ \bibinfo {pages} {713--717} (\bibinfo {year} {2013})}\BibitemShut
  {NoStop}%
\bibitem [{\citenamefont {Kimura}, \citenamefont {Otani},\ and\ \citenamefont
  {Hamrle}(2006)}]{Otani}%
  \BibitemOpen
  \bibfield  {author} {\bibinfo {author} {\bibfnamefont {T.}~\bibnamefont
  {Kimura}}, \bibinfo {author} {\bibfnamefont {Y.}~\bibnamefont {Otani}}, \
  and\ \bibinfo {author} {\bibfnamefont {J.}~\bibnamefont {Hamrle}},\ }\href
  {\doibase 10.1103/PhysRevLett.96.037201} {\bibfield  {journal} {\bibinfo
  {journal} {Phys. Rev. Lett.}\ }\textbf {\bibinfo {volume} {96}},\ \bibinfo
  {pages} {037201} (\bibinfo {year} {2006})}\BibitemShut {NoStop}%
\bibitem [{\citenamefont {Takahashi}\ \emph {et~al.}(2012)\citenamefont
  {Takahashi}, \citenamefont {Kasai}, \citenamefont {Hirayama}, \citenamefont
  {Mitani},\ and\ \citenamefont {Hono}}]{Takahashi}%
  \BibitemOpen
  \bibfield  {author} {\bibinfo {author} {\bibfnamefont {Y.~K.}\ \bibnamefont
  {Takahashi}}, \bibinfo {author} {\bibfnamefont {S.}~\bibnamefont {Kasai}},
  \bibinfo {author} {\bibfnamefont {S.}~\bibnamefont {Hirayama}}, \bibinfo
  {author} {\bibfnamefont {S.}~\bibnamefont {Mitani}}, \ and\ \bibinfo {author}
  {\bibfnamefont {K.}~\bibnamefont {Hono}},\ }\href {\doibase
  http://dx.doi.org/10.1063/1.3681804} {\bibfield  {journal} {\bibinfo
  {journal} {Applied Physics Letters}\ }\textbf {\bibinfo {volume} {100}},\
  \bibinfo {eid} {052405} (\bibinfo {year} {2012})}\BibitemShut {NoStop}%
\bibitem [{\citenamefont {Bridoux}\ \emph {et~al.}(2011)\citenamefont
  {Bridoux}, \citenamefont {Costache}, \citenamefont {Van~de Vondel},
  \citenamefont {Neumann},\ and\ \citenamefont {Valenzuela}}]{Bridoux}%
  \BibitemOpen
  \bibfield  {author} {\bibinfo {author} {\bibfnamefont {G.}~\bibnamefont
  {Bridoux}}, \bibinfo {author} {\bibfnamefont {M.~V.}\ \bibnamefont
  {Costache}}, \bibinfo {author} {\bibfnamefont {J.}~\bibnamefont {Van~de
  Vondel}}, \bibinfo {author} {\bibfnamefont {I.}~\bibnamefont {Neumann}}, \
  and\ \bibinfo {author} {\bibfnamefont {S.~O.}\ \bibnamefont {Valenzuela}},\
  }\href {\doibase http://dx.doi.org/10.1063/1.3635391} {\bibfield  {journal}
  {\bibinfo  {journal} {Applied Physics Letters}\ }\textbf {\bibinfo {volume}
  {99}},\ \bibinfo {eid} {102107} (\bibinfo {year} {2011})}\BibitemShut
  {NoStop}%
\bibitem [{\citenamefont {Hamaya}\ \emph {et~al.}(2012)\citenamefont {Hamaya},
  \citenamefont {Hashimoto}, \citenamefont {Oki}, \citenamefont {Yamada},
  \citenamefont {Miyao},\ and\ \citenamefont {Kimura}}]{Hamaya}%
  \BibitemOpen
  \bibfield  {author} {\bibinfo {author} {\bibfnamefont {K.}~\bibnamefont
  {Hamaya}}, \bibinfo {author} {\bibfnamefont {N.}~\bibnamefont {Hashimoto}},
  \bibinfo {author} {\bibfnamefont {S.}~\bibnamefont {Oki}}, \bibinfo {author}
  {\bibfnamefont {S.}~\bibnamefont {Yamada}}, \bibinfo {author} {\bibfnamefont
  {M.}~\bibnamefont {Miyao}}, \ and\ \bibinfo {author} {\bibfnamefont
  {T.}~\bibnamefont {Kimura}},\ }\href {\doibase 10.1103/PhysRevB.85.100404}
  {\bibfield  {journal} {\bibinfo  {journal} {Phys. Rev. B}\ }\textbf {\bibinfo
  {volume} {85}},\ \bibinfo {pages} {100404} (\bibinfo {year}
  {2012})}\BibitemShut {NoStop}%
\bibitem [{\citenamefont {Kimura}\ \emph {et~al.}(2012)\citenamefont {Kimura},
  \citenamefont {Hashimoto}, \citenamefont {Yamada}, \citenamefont {Miyao},\
  and\ \citenamefont {Hamaya}}]{KimuraCFS}%
  \BibitemOpen
  \bibfield  {author} {\bibinfo {author} {\bibfnamefont {T.}~\bibnamefont
  {Kimura}}, \bibinfo {author} {\bibfnamefont {N.}~\bibnamefont {Hashimoto}},
  \bibinfo {author} {\bibfnamefont {S.}~\bibnamefont {Yamada}}, \bibinfo
  {author} {\bibfnamefont {M.}~\bibnamefont {Miyao}}, \ and\ \bibinfo {author}
  {\bibfnamefont {K.}~\bibnamefont {Hamaya}},\ }\href@noop {} {\bibfield
  {journal} {\bibinfo  {journal} {NPG Asia Materials}\ }\textbf {\bibinfo
  {volume} {4}},\ \bibinfo {eid} {e9} (\bibinfo {year} {2012})}\BibitemShut
  {NoStop}%
\bibitem [{\citenamefont {Yamasaki}\ \emph {et~al.}(2015)\citenamefont
  {Yamasaki}, \citenamefont {Oki}, \citenamefont {Yamada}, \citenamefont
  {Kanashima},\ and\ \citenamefont {Hamaya}}]{Kento}%
  \BibitemOpen
  \bibfield  {author} {\bibinfo {author} {\bibfnamefont {K.}~\bibnamefont
  {Yamasaki}}, \bibinfo {author} {\bibfnamefont {S.}~\bibnamefont {Oki}},
  \bibinfo {author} {\bibfnamefont {S.}~\bibnamefont {Yamada}}, \bibinfo
  {author} {\bibfnamefont {T.}~\bibnamefont {Kanashima}}, \ and\ \bibinfo
  {author} {\bibfnamefont {K.}~\bibnamefont {Hamaya}},\ }\href
  {http://stacks.iop.org/1882-0786/8/i=4/a=043003} {\bibfield  {journal}
  {\bibinfo  {journal} {Applied Physics Express}\ }\textbf {\bibinfo {volume}
  {8}},\ \bibinfo {pages} {043003} (\bibinfo {year} {2015})}\BibitemShut
  {NoStop}%
\bibitem [{\citenamefont {Hu}, \citenamefont {Itoh},\ and\ \citenamefont
  {Kimura}(2014)}]{Hu}%
  \BibitemOpen
  \bibfield  {author} {\bibinfo {author} {\bibfnamefont {S.}~\bibnamefont
  {Hu}}, \bibinfo {author} {\bibfnamefont {H.}~\bibnamefont {Itoh}}, \ and\
  \bibinfo {author} {\bibfnamefont {T.}~\bibnamefont {Kimura}},\ }\href@noop {}
  {\bibfield  {journal} {\bibinfo  {journal} {NPG Asia Materials}\ }\textbf
  {\bibinfo {volume} {6}},\ \bibinfo {eid} {e127} (\bibinfo {year}
  {2014})}\BibitemShut {NoStop}%
\bibitem [{\citenamefont {Yang}\ \emph {et~al.}(2013)\citenamefont {Yang},
  \citenamefont {Kang}, \citenamefont {Chen},\ and\ \citenamefont
  {Xue}}]{Yang2}%
  \BibitemOpen
  \bibfield  {author} {\bibinfo {author} {\bibfnamefont {F.}~\bibnamefont
  {Yang}}, \bibinfo {author} {\bibfnamefont {Z.}~\bibnamefont {Kang}}, \bibinfo
  {author} {\bibfnamefont {X.}~\bibnamefont {Chen}}, \ and\ \bibinfo {author}
  {\bibfnamefont {Y.}~\bibnamefont {Xue}},\ }\href
  {http://stacks.iop.org/0022-3727/46/i=32/a=325003} {\bibfield  {journal}
  {\bibinfo  {journal} {Journal of Physics D: Applied Physics}\ }\textbf
  {\bibinfo {volume} {46}},\ \bibinfo {pages} {325003} (\bibinfo {year}
  {2013})}\BibitemShut {NoStop}%
\bibitem [{\citenamefont {Jourdan}\ \emph {et~al.}(2014)\citenamefont
  {Jourdan}, \citenamefont {Min\'{a}r}, \citenamefont {Kronenberg},
  \citenamefont {Chadov}, \citenamefont {Balke}, \citenamefont {A.},
  \citenamefont {Elmers}, \citenamefont {Sch\"onhense}, \citenamefont {Ebert},
  \citenamefont {Felser},\ and\ \citenamefont {Kl\"aui}}]{Jourdan}%
  \BibitemOpen
  \bibfield  {author} {\bibinfo {author} {\bibfnamefont {M.}~\bibnamefont
  {Jourdan}}, \bibinfo {author} {\bibfnamefont {J.}~\bibnamefont {Min\'{a}r},
  \bibfnamefont {Braun}}, \bibinfo {author} {\bibfnamefont {A.}~\bibnamefont
  {Kronenberg}}, \bibinfo {author} {\bibfnamefont {S.}~\bibnamefont {Chadov}},
  \bibinfo {author} {\bibfnamefont {B.}~\bibnamefont {Balke}}, \bibinfo
  {author} {\bibfnamefont {G.}~\bibnamefont {A.}}, \bibinfo {author}
  {\bibfnamefont {H.~J.}\ \bibnamefont {Elmers}}, \bibinfo {author}
  {\bibfnamefont {G.}~\bibnamefont {Sch\"onhense}}, \bibinfo {author}
  {\bibfnamefont {H.}~\bibnamefont {Ebert}}, \bibinfo {author} {\bibfnamefont
  {C.}~\bibnamefont {Felser}}, \ and\ \bibinfo {author} {\bibfnamefont
  {M.}~\bibnamefont {Kl\"aui}},\ }\href@noop {} {\bibfield  {journal} {\bibinfo
   {journal} {Nature Communications}\ }\textbf {\bibinfo {volume} {5}},\
  \bibinfo {eid} {3974} (\bibinfo {year} {2014})}\BibitemShut {NoStop}%
\bibitem [{\citenamefont {Schlachter}\ \emph {et~al.}(2010)\citenamefont
  {Schlachter}, \citenamefont {Bakker}, \citenamefont {Adam},\ and\
  \citenamefont {van Wees}}]{Schlachter}%
  \BibitemOpen
  \bibfield  {author} {\bibinfo {author} {\bibfnamefont {A.}~\bibnamefont
  {Schlachter}}, \bibinfo {author} {\bibfnamefont {J.~L.}\ \bibnamefont
  {Bakker}}, \bibinfo {author} {\bibfnamefont {J.-P.}\ \bibnamefont {Adam}}, \
  and\ \bibinfo {author} {\bibfnamefont {B.~J.}\ \bibnamefont {van Wees}},\
  }\href@noop {} {\bibfield  {journal} {\bibinfo  {journal} {Nature
  Communications}\ }\textbf {\bibinfo {volume} {6}},\ \bibinfo {pages}
  {879–--882} (\bibinfo {year} {2010})}\BibitemShut {NoStop}%
\bibitem [{\citenamefont {Bakker}\ \emph {et~al.}(2010)\citenamefont {Bakker},
  \citenamefont {Slachter}, \citenamefont {Adam},\ and\ \citenamefont {van
  Wees}}]{Bakker}%
  \BibitemOpen
  \bibfield  {author} {\bibinfo {author} {\bibfnamefont {F.~L.}\ \bibnamefont
  {Bakker}}, \bibinfo {author} {\bibfnamefont {A.}~\bibnamefont {Slachter}},
  \bibinfo {author} {\bibfnamefont {J.-P.}\ \bibnamefont {Adam}}, \ and\
  \bibinfo {author} {\bibfnamefont {B.~J.}\ \bibnamefont {van Wees}},\ }\href
  {\doibase 10.1103/PhysRevLett.105.136601} {\bibfield  {journal} {\bibinfo
  {journal} {Phys. Rev. Lett.}\ }\textbf {\bibinfo {volume} {105}},\ \bibinfo
  {pages} {136601} (\bibinfo {year} {2010})}\BibitemShut {NoStop}%
\bibitem [{\citenamefont {Erekhinsky}\ \emph {et~al.}(2012)\citenamefont
  {Erekhinsky}, \citenamefont {Casanova}, \citenamefont {Schuller},\ and\
  \citenamefont {Sharoni}}]{Erekhinsky}%
  \BibitemOpen
  \bibfield  {author} {\bibinfo {author} {\bibfnamefont {M.}~\bibnamefont
  {Erekhinsky}}, \bibinfo {author} {\bibfnamefont {F.}~\bibnamefont
  {Casanova}}, \bibinfo {author} {\bibfnamefont {I.~K.}\ \bibnamefont
  {Schuller}}, \ and\ \bibinfo {author} {\bibfnamefont {A.}~\bibnamefont
  {Sharoni}},\ }\href@noop {} {\bibfield  {journal} {\bibinfo  {journal} {Appl.
  Phys. Lett}\ }\textbf {\bibinfo {volume} {100}} (\bibinfo {year}
  {2012})}\BibitemShut {NoStop}%
\bibitem [{\citenamefont {Kimura}\ \emph {et~al.}(2004)\citenamefont {Kimura},
  \citenamefont {Hamrle}, \citenamefont {Otani}, \citenamefont {Tsukagoshi},\
  and\ \citenamefont {Aoyagi}}]{Kimura3}%
  \BibitemOpen
  \bibfield  {author} {\bibinfo {author} {\bibfnamefont {T.}~\bibnamefont
  {Kimura}}, \bibinfo {author} {\bibfnamefont {J.}~\bibnamefont {Hamrle}},
  \bibinfo {author} {\bibfnamefont {Y.}~\bibnamefont {Otani}}, \bibinfo
  {author} {\bibfnamefont {K.}~\bibnamefont {Tsukagoshi}}, \ and\ \bibinfo
  {author} {\bibfnamefont {Y.}~\bibnamefont {Aoyagi}},\ }\href@noop {}
  {\bibfield  {journal} {\bibinfo  {journal} {Appl. Phys. Lett.}\ }\textbf
  {\bibinfo {volume} {85}},\ \bibinfo {pages} {3795--3796} (\bibinfo {year}
  {2004})}\BibitemShut {NoStop}%
\bibitem [{\citenamefont {Kimura}, \citenamefont {Sato},\ and\ \citenamefont
  {Otani}(2008{\natexlab{a}})}]{Kimura2}%
  \BibitemOpen
  \bibfield  {author} {\bibinfo {author} {\bibfnamefont {T.}~\bibnamefont
  {Kimura}}, \bibinfo {author} {\bibfnamefont {T.}~\bibnamefont {Sato}}, \ and\
  \bibinfo {author} {\bibfnamefont {Y.}~\bibnamefont {Otani}},\ }\href
  {\doibase 10.1103/PhysRevLett.100.066602} {\bibfield  {journal} {\bibinfo
  {journal} {Phys. Rev. Lett.}\ }\textbf {\bibinfo {volume} {100}},\ \bibinfo
  {pages} {066602} (\bibinfo {year} {2008}{\natexlab{a}})}\BibitemShut
  {NoStop}%
\bibitem [{\citenamefont {Villamor}\ \emph {et~al.}(2013)\citenamefont
  {Villamor}, \citenamefont {Isasa}, \citenamefont {Hueso},\ and\ \citenamefont
  {Casanova}}]{Villamor}%
  \BibitemOpen
  \bibfield  {author} {\bibinfo {author} {\bibfnamefont {E.}~\bibnamefont
  {Villamor}}, \bibinfo {author} {\bibfnamefont {M.}~\bibnamefont {Isasa}},
  \bibinfo {author} {\bibfnamefont {L.~E.}\ \bibnamefont {Hueso}}, \ and\
  \bibinfo {author} {\bibfnamefont {F.}~\bibnamefont {Casanova}},\ }\href@noop
  {} {\bibfield  {journal} {\bibinfo  {journal} {Phys. Rev. B.}\ }\textbf
  {\bibinfo {volume} {87}},\ \bibinfo {pages} {094417} (\bibinfo {year}
  {2013})}\BibitemShut {NoStop}%
\bibitem [{\citenamefont {O'Brien}\ \emph {et~al.}(2014)\citenamefont
  {O'Brien}, \citenamefont {Erickson}, \citenamefont {Spivak}, \citenamefont
  {Ambaye}, \citenamefont {Goyette}, \citenamefont {Lauter}, \citenamefont
  {Crowel},\ and\ \citenamefont {Leighton}}]{OBrien}%
  \BibitemOpen
  \bibfield  {author} {\bibinfo {author} {\bibfnamefont {L.}~\bibnamefont
  {O'Brien}}, \bibinfo {author} {\bibfnamefont {M.~J.}\ \bibnamefont
  {Erickson}}, \bibinfo {author} {\bibfnamefont {D.}~\bibnamefont {Spivak}},
  \bibinfo {author} {\bibfnamefont {H.}~\bibnamefont {Ambaye}}, \bibinfo
  {author} {\bibfnamefont {R.~J.}\ \bibnamefont {Goyette}}, \bibinfo {author}
  {\bibfnamefont {V.}~\bibnamefont {Lauter}}, \bibinfo {author} {\bibfnamefont
  {P.~A.}\ \bibnamefont {Crowel}}, \ and\ \bibinfo {author} {\bibfnamefont
  {C.}~\bibnamefont {Leighton}},\ }\href@noop {} {\bibfield  {journal}
  {\bibinfo  {journal} {Nature Communications}\ }\textbf {\bibinfo {volume}
  {5}},\ \bibinfo {eid} {3927} (\bibinfo {year} {2014})}\BibitemShut {NoStop}%
\bibitem [{\citenamefont {Motzko}\ \emph {et~al.}(2014)\citenamefont {Motzko},
  \citenamefont {Richter}, \citenamefont {Burkhardt}, \citenamefont {Reeve},
  \citenamefont {Laczkowski}, \citenamefont {Vila}, \citenamefont
  {Attan\'{e}},\ and\ \citenamefont {Kl\"aui}}]{Motzko2}%
  \BibitemOpen
  \bibfield  {author} {\bibinfo {author} {\bibfnamefont {N.}~\bibnamefont
  {Motzko}}, \bibinfo {author} {\bibfnamefont {N.}~\bibnamefont {Richter}},
  \bibinfo {author} {\bibfnamefont {B.}~\bibnamefont {Burkhardt}}, \bibinfo
  {author} {\bibfnamefont {R.}~\bibnamefont {Reeve}}, \bibinfo {author}
  {\bibfnamefont {P.}~\bibnamefont {Laczkowski}}, \bibinfo {author}
  {\bibfnamefont {L.}~\bibnamefont {Vila}}, \bibinfo {author} {\bibfnamefont
  {J.-P.}\ \bibnamefont {Attan\'{e}}}, \ and\ \bibinfo {author} {\bibfnamefont
  {M.}~\bibnamefont {Kl\"aui}},\ }\href {\doibase 10.1002/pssa.201300695}
  {\bibfield  {journal} {\bibinfo  {journal} {physica status solidi (a)}\
  }\textbf {\bibinfo {volume} {211}},\ \bibinfo {pages} {986--990} (\bibinfo
  {year} {2014})}\BibitemShut {NoStop}%
\bibitem [{\citenamefont {Takahashi}\ and\ \citenamefont
  {Maekawa}(2003)}]{MaekawaTaka}%
  \BibitemOpen
  \bibfield  {author} {\bibinfo {author} {\bibfnamefont {S.}~\bibnamefont
  {Takahashi}}\ and\ \bibinfo {author} {\bibfnamefont {S.}~\bibnamefont
  {Maekawa}},\ }\href {\doibase 10.1103/PhysRevB.67.052409} {\bibfield
  {journal} {\bibinfo  {journal} {Phys. Rev. B}\ }\textbf {\bibinfo {volume}
  {67}},\ \bibinfo {pages} {052409} (\bibinfo {year} {2003})}\BibitemShut
  {NoStop}%
\bibitem [{\citenamefont {Kimura}, \citenamefont {Sato},\ and\ \citenamefont
  {Otani}(2008{\natexlab{b}})}]{Kimura}%
  \BibitemOpen
  \bibfield  {author} {\bibinfo {author} {\bibfnamefont {T.}~\bibnamefont
  {Kimura}}, \bibinfo {author} {\bibfnamefont {T.}~\bibnamefont {Sato}}, \ and\
  \bibinfo {author} {\bibfnamefont {Y.}~\bibnamefont {Otani}},\ }\href
  {\doibase 10.1103/PhysRevLett.100.066602} {\bibfield  {journal} {\bibinfo
  {journal} {Phys. Rev. Lett.}\ }\textbf {\bibinfo {volume} {100}},\ \bibinfo
  {pages} {066602} (\bibinfo {year} {2008}{\natexlab{b}})}\BibitemShut
  {NoStop}%
\bibitem [{\citenamefont {H.}\ and\ \citenamefont {Ji}(2012)}]{Zou}%
  \BibitemOpen
  \bibfield  {author} {\bibinfo {author} {\bibfnamefont {Z.}~\bibnamefont
  {H.}}\ and\ \bibinfo {author} {\bibfnamefont {Y.}~\bibnamefont {Ji}},\
  }\href@noop {} {\bibfield  {journal} {\bibinfo  {journal} {Appl. Phys.
  Lett.}\ }\textbf {\bibinfo {volume} {101}},\ \bibinfo {pages} {082401}
  (\bibinfo {year} {2012})}\BibitemShut {NoStop}%
\bibitem [{\citenamefont {Casanova}\ \emph {et~al.}(2009)\citenamefont
  {Casanova}, \citenamefont {Sharoni}, \citenamefont {Erekhinsky},\ and\
  \citenamefont {Schuller}}]{Casanova}%
  \BibitemOpen
  \bibfield  {author} {\bibinfo {author} {\bibfnamefont {F.}~\bibnamefont
  {Casanova}}, \bibinfo {author} {\bibfnamefont {A.}~\bibnamefont {Sharoni}},
  \bibinfo {author} {\bibfnamefont {M.}~\bibnamefont {Erekhinsky}}, \ and\
  \bibinfo {author} {\bibfnamefont {I.~K.}\ \bibnamefont {Schuller}},\
  }\href@noop {} {\bibfield  {journal} {\bibinfo  {journal} {Phys. Rev. B.}\
  }\textbf {\bibinfo {volume} {79}},\ \bibinfo {pages} {184415} (\bibinfo
  {year} {2009})}\BibitemShut {NoStop}%
\bibitem [{\citenamefont {Erekhinsky}\ \emph {et~al.}(2010)\citenamefont
  {Erekhinsky}, \citenamefont {Sharoni}, \citenamefont {Casanova},\ and\
  \citenamefont {Schuller}}]{Erekhinsky2}%
  \BibitemOpen
  \bibfield  {author} {\bibinfo {author} {\bibfnamefont {M.}~\bibnamefont
  {Erekhinsky}}, \bibinfo {author} {\bibfnamefont {A.}~\bibnamefont {Sharoni}},
  \bibinfo {author} {\bibfnamefont {F.}~\bibnamefont {Casanova}}, \ and\
  \bibinfo {author} {\bibfnamefont {I.~K.}\ \bibnamefont {Schuller}},\ }\href
  {\doibase http://dx.doi.org/10.1063/1.3291047} {\bibfield  {journal}
  {\bibinfo  {journal} {Applied Physics Letters}\ }\textbf {\bibinfo {volume}
  {96}},\ \bibinfo {eid} {022513} (\bibinfo {year} {2010})}\BibitemShut
  {NoStop}%
\bibitem [{\citenamefont {Brass}\ and\ \citenamefont
  {Pratt~Jr.}(2006)}]{Brass}%
  \BibitemOpen
  \bibfield  {author} {\bibinfo {author} {\bibfnamefont {J.}~\bibnamefont
  {Brass}}\ and\ \bibinfo {author} {\bibfnamefont {W.~P.}\ \bibnamefont
  {Pratt~Jr.}},\ }\href@noop {} {\bibfield  {journal} {\bibinfo  {journal}
  {Journal of Physics Condensed Matter}\ }\textbf {\bibinfo {volume} {19}},\
  \bibinfo {eid} {183201} (\bibinfo {year} {2006})}\BibitemShut {NoStop}%
\bibitem [{\citenamefont {Elliott}(1954)}]{Elliott}%
  \BibitemOpen
  \bibfield  {author} {\bibinfo {author} {\bibfnamefont {R.~J.}\ \bibnamefont
  {Elliott}},\ }\href {http://link.aps.org/doi/10.1103/PhysRev.96.266}
  {\bibfield  {journal} {\bibinfo  {journal} {Phys. Rev.}\ }\textbf {\bibinfo
  {volume} {96}},\ \bibinfo {pages} {266--279} (\bibinfo {year}
  {1954})}\BibitemShut {NoStop}%
\end{thebibliography}%

\end{document}